\documentclass[10pt,a4paper]{article}

\usepackage[left=20mm, top=20mm, right=20mm, bottom=20mm, nohead, foot=10mm]{geometry}

\usepackage[utf8]{inputenc}
\usepackage[english]{babel}

\usepackage[affil-it]{authblk}

\usepackage{multicol}

\usepackage{caption}

\usepackage{amsmath,amssymb}

\usepackage{xcolor}
\usepackage[normalem]{ulem}

\usepackage{titlesec}
\titleformat{\section}[block]{\normalfont\sffamily}{}{.5em}{\bfseries}
\titleformat{\subsection}[block]{\normalfont\sffamily}{}{.5em}{\bfseries}

\usepackage{hyperref}
\hypersetup{colorlinks=true, linkcolor=[rgb]{0.6 0 0}, citecolor=[rgb]{0 0.4 0}, urlcolor=[rgb]{0 0 0.6}}

\usepackage{graphicx}

\newenvironment{FigureOneColumn}
  {\par\medskip\noindent\minipage{\linewidth}\captionsetup{type=figure}}
  {\endminipage\par\medskip}

\usepackage{csquotes}
\usepackage[backend=biber, style=phys, doi=false, url=false, hyperref=true]{biblatex}
\begin{document}

\title{\sffamily Antichiral Ferromagnetism}

\author[1]{\normalsize Filipp N. Rybakov\thanks{f.n.rybakov@gmail.com}}
\author[2]{\normalsize Anastasiia Pervishko}
\author[3]{\normalsize Olle Eriksson}
\author[1]{\normalsize Egor Babaev}

\affil[1]{\small
Department of Physics, KTH Royal Institute of Technology, SE-10691 Stockholm, Sweden
}

\affil[2]{\small
Skolkovo Institute of Science and Technology, Moscow 121205, Russia
}

\affil[3]{\small
Department of Physics and Astronomy, Uppsala University, Box 516, SE-75120 Uppsala, Sweden
}

\date{}

\clubpenalty=10000
\widowpenalty=10000

\maketitle
\vspace{-3.0\baselineskip}

\renewcommand{\abstractname}{\vspace{-\baselineskip}}
\begin{abstract}

Here by combining a symmetry-based analysis with numerical computations we predict a new kind of magnetic ordering -- \textit{antichiral ferromagnetism}. 
The relationship between chiral and antichiral magnetic order is conceptually similar to the relationship between ferromagnetic and antiferromagnetic order.
Without loss of generality, we focus our investigation on crystals with full tetrahedral symmetry where chiral interaction terms -- Lifshitz invariants -- are forbidden by symmetry. However, we demonstrate that leading chirality-related term 
leads to nontrivial smooth magnetic textures in the form of helix-like segments of alternating opposite chiralities.
The unconventional order manifests itself beyond the ground state by stabilizing excitations such as domains and skyrmions in an antichiral form.

\end{abstract}

\begin{multicols}{2}[]

\begin{refsegment}
 
Chirality and chiral textures is a cornerstone concept in many fields of physics ranging from cosmology to nuclear and elementary particle physics~\cite{kharzeev2014chiral}.
In many materials, chiral crystal structure results in a chiral ferromagnetic ordering~\cite{Bak_1980,Tokunaga2015}.
The chiral textures in  magnetism attracted renewed interest in recent decades thanks to their potential relevance to technological applications, including alternative logic devices and racetrack memory where the information is encoded by virtue of magnetic  textures~\cite{Parkin2008,Fert2013,Koumpouras2016,Back_2020}.
Interestingly, some of the concepts developed so far are intended for the use of materials that have ferromagnetic ordering at the atomistic level, while at the mesoscale the direction of magnetization $\mathbf{m}(\mathbf{r})$ is modulated.
The corresponding basic types of modulations are one-dimensional and conventionally called spirals, helices, cycloids, screws, to name a few~\cite{Izyumov_1984}.

The emergence of modulated ferromagnetic ordering may appear due to competing symmetric interactions between atoms in a lattice~\cite{VILLAIN1959303, 1959807}. The multi-spin exchange interactions, in particular four-spin interactions, also provide a possible mechanism for the modulated texture stabilization beyond pairwise scenario~\cite{Paul2019}. Most naturally, multi-spin exchange interactions can emerge in the form of two-site biquadratic or four-site spin interaction~\cite{Takahashi1977,MacDonald1988}. The delicate interplay between competing pairs and/or accounting for multi-spin interactions, leading to the stabilization of different magnetic textures, has been investigated both theoretically and experimentally in Refs.~\cite{Kurz2001, Heinze2011,Leonov2015,Hayami2017,Romming2018,Meyer2019}. By virtue of symmetry of the above mentioned and other symmetric interactions, the reflection of modulated magnetic texture is defined by degenerate energy states with opposite chirality (handedness), where a certain chirality is a consequence of spontaneous symmetry breaking.

An alternative mechanism of nucleation and stability of modulated ferromagnetic textures may be attributed to the presence of the pairwise Dzyaloshinskii-Moriya interaction (DMI), which is responsible for an asymmetric nearest-neighbor spin exchange, often discussed in magnetic systems where inversion symmetry is absent. Along with the magnetic ultrathin films and multilayers where the inversion symmetry is broken by natural means~\cite{Fert2013}, the effect of DMI is pronounced in cubic crystals with chiral point group symmetry $T$, such as B20-type FeGe and MnSi, in which a certain chirality of a magnetic helix is dictated by one of two possible enantiomorphic forms of the compound~\cite{Stishov_2011}. In this case, the corresponding magnetic Hamiltonian includes the Lifshitz invariants~\cite{Bar_Stef_1969, Bak_1980}, and the chirality depends on the sign of their common factor.

In this paper, we predict a ferromagnetic ordering that is fundamentally different from those discussed above which we refer to as \textit{antichiral ferromagnetism}. This term aims to reflect that spontaneous modulation of the magnetization direction appears in a way that both types of chirality exist simultaneously, and alternate in space. Our analysis reveal that this magnetic ordering naturally appears in a bulk ferromagnet with the point group symmetry $T_\text{d}$ owing to achiral crystal symmetry. 
This is a class of crystals in which many minerals are formed naturally~\cite{SPORTOUCH2001158,10.2138/am-2019-6856}.

The macroscopic robustness of a magnetic configuration is purely determined by its stability with respect to perturbations that violate spatial uniformity.
If, for instance, inversion symmetry is broken the spin alignment might gain a certain chirality due to spin-orbit driven antisymmetric DMI that contributes to the total energy with the terms linear with respect to the first spatial derivatives of magnetization. In general, the derivative linear contribution to the free energy of a ferromagnet can be casted in the form
\begin{equation}\label{eq:linear}
    \mathcal{H}_\nabla=\int d\mathbf{r}\sum\limits_{\alpha\beta}\Omega_{\alpha\beta}(\mathbf{m})\nabla_\alpha m_\beta,
\end{equation}
with the tensor $\Omega_{\alpha\beta}(-\mathbf{m})=-\Omega_{\alpha\beta}(\mathbf{m})$ being odd under magnetization inversion~\cite{Ado2020}, while $\alpha$, $\beta$ correspond to the spatial indexes. 
By expanding $\Omega_{\alpha\beta}(\mathbf{m})=\sum_\gamma\Omega^{(1)}_{\alpha\beta\gamma}m_\gamma+\sum_{\gamma\delta\varepsilon}\Omega_{\alpha\beta\gamma\delta\varepsilon}^{(3)}m_\gamma m_\delta m_\varepsilon$ in powers of $\mathbf{m}$ and restricting to linear and cubic contribution only one arrives at
\postdisplaypenalty=0
\begin{align}\nonumber
    \mathcal{H}_\nabla&=\frac{1}{2}\int d\mathbf{r}\sum\limits_{\alpha\beta\gamma}\left(\mathcal{D}_{\alpha\beta\gamma}^S\nabla_\alpha(m_\beta m_\gamma)+\mathcal{D}^A_{\alpha\beta\gamma}\mathcal{L}_{\gamma\beta}^{(\alpha)}\right) \\ \label{eq:genenergy}
    &+\int d\mathbf{r}\sum\limits_{\alpha\beta\gamma\delta}\Omega^{(3)}_{\alpha\beta\gamma\delta\varepsilon}m_\gamma m_\delta m_\varepsilon\nabla_\alpha m_\beta,
\end{align}
where for convenience we have separated the terms arising due to $\Omega^{(1)}$ into two parts, namely are symmetric terms, $\nabla_\alpha(m_\beta m_\gamma)$, and terms given in the form of Lifshitz invariants, $\mathcal{L}_{\beta\gamma}^{(\alpha)}=m_\beta\nabla_\alpha m_\gamma-m_\gamma\nabla_\alpha m_\beta$~\cite{Bogdanov1989}. Note that as the symmetric contribution can be expressed in terms of surface integrals via Stokes' theorem~\cite{Ado2020} its impact on the magnetic ordering in bulk crystals can be discarded.

To proceed, we analyze~(\ref{eq:genenergy}) based on symmetry grounds (see the Appendices). The solution for $T_\text{d}$ point group symmetry is trivial with respect to Lifshitz invariants, $\mathcal{D}^A_{\alpha\beta\gamma}=0$, whereas in the case of  $\Omega^{(3)}$ one can identify four independent components with the corresponding invariants given by 
\postdisplaypenalty=0
\begin{subequations}
\begin{gather}
\sum\nolimits_{\prime}\nabla_\alpha(m_\beta^3 m_\gamma + m_\beta m_\gamma^3), \ 
\sum\nolimits_{\prime}\nabla_\alpha(m_\alpha^2 m_\beta m_\gamma), \label{eq:invA}\\
\sum\nolimits_{\prime}{m_\alpha m_\beta \nabla_\gamma (\mathbf{m}^2)}, \label{eq:invB}\\
m_x m_y m_z\nabla\cdot\mathbf{m}, \label{eq:invC}
\end{gather}
\end{subequations}
where ${\sum\nolimits_{\prime}}$ denote the sum over $(\alpha,\beta,\gamma)\in\{(x,y,z),(y,z,x),(z,x,y)\}$. 
The invariant~(\ref{eq:invC}) was first reported by I.~Ado~ \textit{et al}.~\cite{Ado_SPICE_2020,Titov_TCM_2020}, whereas we derive the complete set (\ref{eq:invA})-(\ref{eq:invC}).  
After integrating by parts both terms~(\ref{eq:invA}) can be discarded for the same reasons as the terms $\nabla_\alpha(m_\beta m_\gamma)$ in the above analysis. Note that within the micromagnetic approach magnetization is described in terms of a unit vector ${|\mathbf{m}(\mathbf{r})|=1}$ and thus the invariant ~(\ref{eq:invB}) vanishes as well. Therefore we end up with ~(\ref{eq:invC}) as the only relevant term for Eq.~(\ref{eq:linear}).

We base our subsequent analysis on the minimal model of a tetrahedral ferromagnet that contains exchange interaction and Ado interaction (\ref{eq:invC}):
\begin{equation}\label{functional}
    \mathcal{H}=\int d\mathbf{r}\left(A\vert \nabla \mathbf{m}\vert^2+\mathcal{B}m_xm_ym_z\nabla\cdot\mathbf{m}\right),
\end{equation}
where $A$ is the exchange stiffness. It worth noting that in a somewhat more general case, the energy density in~(\ref{functional}) may be equipped with a magnetic-field-induced Zeeman term, cubic anisotropy $\propto\left(m_x^2m_y^2+m_x^2m_z^2+m_y^2m_z^2\right)$, and extensions in the case of exchange anisotropy. Without loss of generality, we assume ${\mathcal{B}>0}$, since the sign of this constant depends on the choice of the coordinates, see Fig.~\ref{fig:figure1}.
To identify the lowest energy state, as well as stable excited configurations, we perform a numerical minimization of the energy (see the Appendices).

\begin{FigureOneColumn}
    \centering
    \includegraphics[width=7.2cm]{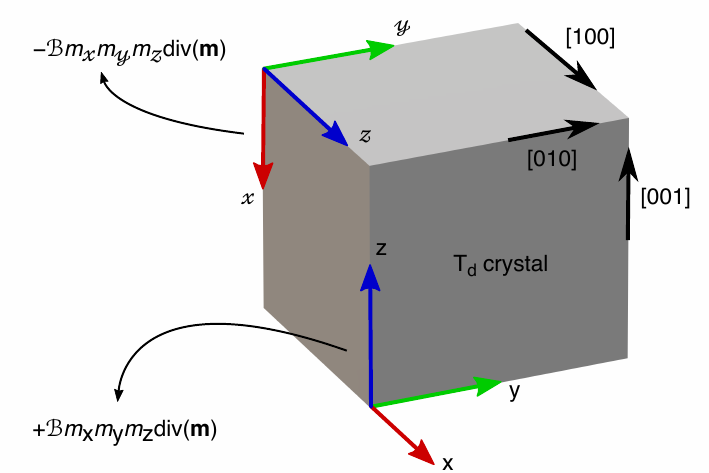}
    \caption{
\textbf{A sketch of bulk crystal with achiral point group symmetry $T_\text{d}$.} 
By virtue of rotoreflection $S_4$ invariants under the crystallographic point group $T_\text{d}$, the sign of the phenomenological constant $\mathcal{B}$ alternates at 90$^{\circ}$ rotations of the coordinate frame.
}
    \label{fig:figure1}
\end{FigureOneColumn}

\begin{figure*}[t]
    \centering
    \includegraphics[width=14.4cm]{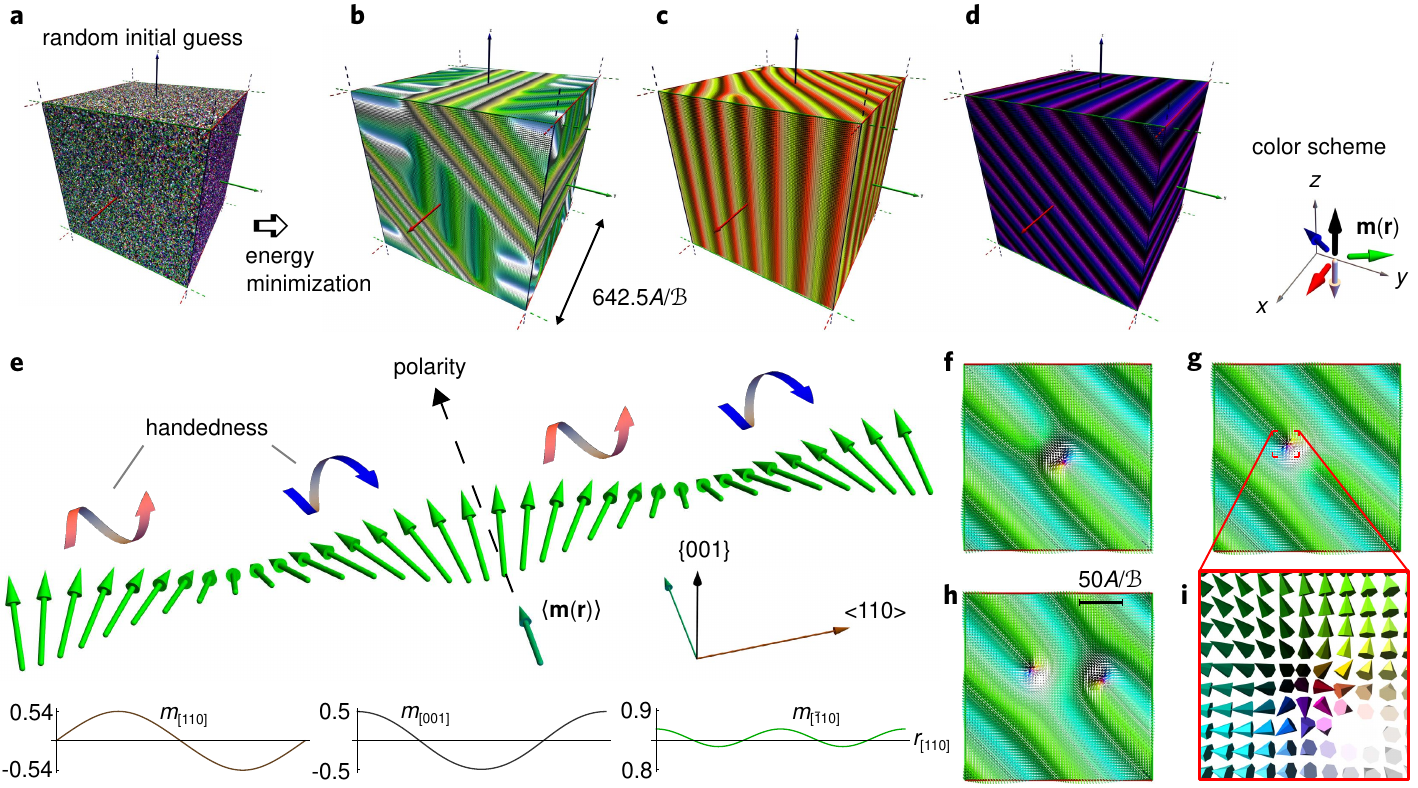}
    \caption{
\textbf{Results of numerical simulations obtained by energy minimization.}  
\textbf{a}, Initial guess with chaotically oriented magnetization for the cube under periodic boundary conditions. 
\textbf{b} -- \textbf{d}, Typical solutions corresponding to an energy minimum: multi-domain structure, magnetization modulations with edge dislocations, and perfect helix-like modulations, respectively.
\textbf{e}, Antichiral magnetic texture that corresponds to the global energy minimum. The optimal period, $L$ is found to be~${\approx75.72A/\mathcal{B}}$, while the net magnetization is ${|\langle \mathbf{m}(\mathbf{r}) \rangle|\approx 0.854}$. The figure captures two periods, while the inserts at the bottom -- one period.
\textbf{f} -- \textbf{h}, Stable antichiral excitations, including \textbf{f}, antiskyrmion, \textbf{g}, skyrmion, and \textbf{h}, a cluster composed by a pair of skyrmion and antiskyrmion. The approximate energy values of these states are 12.02, 12.02, 23.58, respectively. The energy is calculated relative to the global minimum in units of ${2 A t}$, where $t$ is the thickness in the direction perpendicular to the picture plane. 
\textbf{i}, Enlarged area of the skyrmion core.
}
    \label{fig:figure2}
\end{figure*}

In Figs.~\ref{fig:figure2}(b)-(d) we provide relevant energy minimum configurations that emerge from an initial guess with randomly oriented spins, as depicted in Fig.~\ref{fig:figure2}(a). We observe that the numerical solutions to (\ref{functional}) can be roughly classified into three different groups, namely are multi-domain textures shown in Fig.~\ref{fig:figure2}(b), modulated states with small inclusions such as edge dislocations illustrated in Fig.~\ref{fig:figure2}(c), and defect-free periodic modulations unveiled in Fig.~\ref{fig:figure2}(d). Interestingly, the magnetic structure in Fig.~\ref{fig:figure2}(d) was found to be the energetically most favorable. Reducing the spatial dimension to two, while increasing the size of the modeling area, we also obtained that the type of modulations depicted in Fig.~\ref{fig:figure2}(d) is the most energetically favourable configuration.
A closer inspection of the magnetic ordering in the structure is visualized in Fig.~\ref{fig:figure2}(e). It can be clearly seen that locally in space the magnetic spiral has a spontaneous distinct chirality. However the sign of the chirality varies in space, resulting in a bichiral configuration that being averaged over a modulation period produces zero. It is therefore not surprising that the relationship between the local chirality as depicted in Fig.~\ref{fig:figure2}(e) is in a way similar to the relationship between ferromagnetic and antiferromagnetic ordering of individual atomic moments. Here, alternating local moments are effectively replaced with alternating local chiralities. On this account we label this state as \textit{antichiral}. 
The polarity demonstrated in Fig.~\ref{fig:figure2}(e) is twofold degenerate.
The spatial orientation of the texture is sixfold degenerate, and the resulting twelvefold degeneracy should be spontaneously broken in this model. In a finite sample, we expect that the significant net magnetization and stray fields will favor the split into the domains.

Having established the ground state of the system we proceed with the discussion on the possibility of stable excited configurations, apart from those found by numerical energy minimization from the random state. Notably, several classes of magnets allow particle-like topological excitations, such as magnetic skyrmions~\cite{Bogdanov1989, Fert2013, Back_2020}. In order to resolve numerically whether the present system has similar type excitations we use the vortex-like ansatz~\cite{Bogdanov1989} as an initial guess, superimposing modulation mimicking the ground state discussed above. Our findings suggest that the model possesses stable states in shape of skyrmions that form over antichiral background. Some of the solutions with distinct topology that have been discovered as a result of numerical energy minimization are shown in Fig.~\ref{fig:figure2}(f)-(h). Interestingly, contrasting to their counterparts in chiral magnets, skyrmions that we obtain in this model do not have a predefined chirality.

We demonstrated that the magnetic ordering in crystals with $T_\text{d}$ (tetrahedral class) point group can be of a novel type, that we refer to as \textit{antichiral ferromagnetism}. 
Among the methods that may be used to probe the predicted \textit{antichiral} ordering one can highlight small-angle neutron scattering, spin-polarized scanning probe microscopy \cite{Crommie1993} as well as the electron magnetic circular dichroism technique \cite{Schattschneider2006}.

Our numerical findings reveal that antichiral ground state has a set of stable topologically non-trivial excitations in the form of edge dislocations and skyrmions. These excitations are antichiral in contrast to their chiral counterparts in systems with Dzyaloshinskii-Moriya interaction. We strongly believe their properties pose an intriguing question for follow-up studies and will trigger experimental activity in this field.

\begin{FigureOneColumn}
    \centering
    \includegraphics[width=7.2cm]{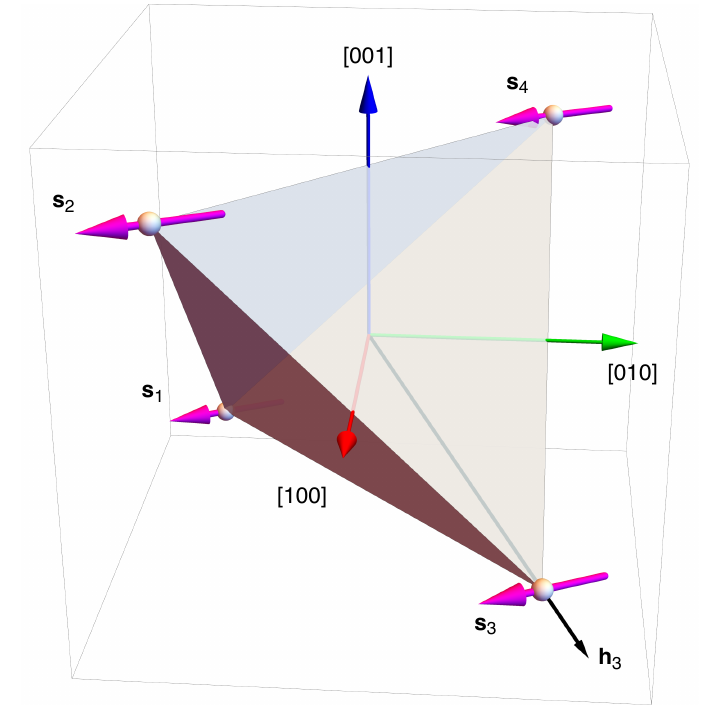}
    \caption{
\textbf{A possible structural building block, satisfying full tetrahedral symmetry.}
 Magnetic spins (arrows), and relevant magnetic interaction paths (see explanation in the text) of the magnetic unit cell. 
    }
    \label{fig:figure3}
\end{FigureOneColumn}

Another interesting aspect of our numerical findings quests for the microscopic origin of the term~(\ref{eq:invC}). Due to anisotropy of the crystal, shown in Fig.~\ref{fig:figure1}, the microscopic Hamiltonian for classical spins $\mathbf{s}_i$ must be equipped with direction vectors. The possible term that corresponds to such a contribution to the micromagnetic energy may be cast in the form
\begin{equation}
H=
\beta\left( \sum_{i=1}^4   \mathbf{s}_i \cdot \mathbf{h}_i \right)
\sum_{i=1}^4  (\mathbf{s}_i \cdot \mathbf{h}_i)^3, \label{microscopicH}
\end{equation}
where unit vectors $\mathbf{h}_i$ are aligned with the bonds connecting the center of a tetrahedron and the corresponding vertex, see Fig.~\ref{fig:figure3}, while $\beta$ represents the strength of the interaction. Note that multi-spin interactions whose Hamiltonian contains direction vectors were recently discovered in B20-type cubic chiral magnets~\cite{Grytsiuk2020}. Moreover, we expect that a certain insight on the nature of pairwise and multi-spin interactions~\cite{Cardias2020, PhysRevB.101.174401} can be gained on the basis of first-principles calculations that could clarify the  microscopic origin of the magnetic interactions leading to antichiral ferromagnetism in $T_\text{d}$ crystals and beyond.

\textit{Note added. } 
Antichiral magnetic ordering presented here should not be confused with the ``helix'' reported earlier in~\cite{Titov_TCM_2020}.
Right after the first version of this manuscript~\cite{2012.05835v1}, I.~A.~Ado \textit{et al}. reported an approximation of the ground state solution of Hamiltonian~(\ref{functional})~\cite{2012.07666v1}.
This approximation is close to our solution and captures antichirality. 
Based on our numerical solution, we suggest a simple and very accurate analytical approximation for the ground state, 
\postdisplaypenalty=0
\begin{subequations}
\begin{gather}
m_{[110]}\approx 0.54 \sin(k\,r_{[110]}), \ 
m_{[001]}\approx 0.5 \cos(k\,r_{[110]}), \nonumber\\
m_{[\bar{1}10]} = \sqrt{ 1 - m_{[110]}^2 - m_{[001]}^2},\nonumber\\ 
k \equiv \frac{2\pi}{L}, \ r_{[110]} \equiv \frac{x+y}{\sqrt{2}}, \ L\approx 76\frac{A}{\mathcal{B}}. \nonumber
\end{gather}
\end{subequations}
This expression gives average energy density  ${\langle\rho_{\mathcal{H}}\rangle\approx -1.8676\cdot10^{-3} \mathcal{B}^2/A}$, while our original numerical result predicts 
\postdisplaypenalty=0
\begin{subequations}
\begin{align}
&\langle\rho_{\mathcal{H}}\rangle = -1.8696582776\cdot10^{-3} \mathcal{B}^2/A,
\nonumber\\
& L = 75.725021\ A/\mathcal{B},\nonumber\\
& |\langle \mathbf{m}(\mathbf{r}) \rangle| = 0.853861965,\nonumber
\end{align}
\end{subequations}
where all digits are expected to be accurate.

The authors of~\cite{2012.07666v1} also proposed an alternative to the version of the microscopic Hamiltonian used here~(\ref{microscopicH}).

\section{Appendices}

\subsection{A1. Symmetry analysis method}

Given a point group symmetry with generators $\mathcal{R}^{(\mu)}$, a tensorial structure is dictated by the symmetry relations
\begin{equation}\label{eq:symmetry}
    \Omega^{(n)}_{\alpha_1\alpha_2\ldots\alpha_n}=\mathcal{R}_{\alpha_1\beta_1}^{(\mu)}\mathcal{R}_{\alpha_2\beta_2}^{(\mu)}\ldots\mathcal{R}_{\alpha_n\beta_n}^{(\mu)}\Omega^{(n)}_{\beta_1\beta_2\ldots\beta_n},
\end{equation}
where $\mathcal{R}_{\alpha\beta}^{(\mu)}$ are orthogonal matrices of three-dimensional irreducible representations for each element $\mu$ of the group. 
The corresponding matrices for the tetrahedral point group $T_\text{d}$ can be found, e.g., in Refs.~\cite{AtkinsFriedman,Xu_Chen_arxiv_2019,Arovas}.
Reduction of the linear system~(\ref{eq:symmetry}) identifies zero and non-zero components of the tensor $\Omega$.

\subsection{A2. Numerical energy minimization}

The continuous model as yielded by Eq.~(\ref{functional}) was discretized using a rectangular grid under periodic boundary conditions. We implemented a discretization scheme giving eighth-order of accuracy by generalizing the approach of Donahue and McMichael~\cite{DONAHUE1997272}. The typical number of grid points in each dimension ranged from 140 to 160. For minimization we used a GPU-parallelized nonlinear conjugate gradient algorithm~\cite{Excalibur}, while the constraint ${|\mathbf{m}|=1}$ was satisfied by means of the special use of stereographic projections (see e.g. Supplemental Material in Ref.~\cite{PhysRevLett.115.117201}).

\section{Acknowledgments}

The work of A.P. was supported by the Russian Science Foundation Project No. 20-72-00044. E.B. and F.N.R. were supported by the Swedish Research Council Grants No. 642-2013-7837, 2016-06122, 2018-03659, and G\"{o}ran Gustafsson Foundation for Research in Natural Sciences. O.E. acknowledges support from the Swedish Research Council, the Knut and Alice Wallenberg foundation, the Foundation for Strategic Research, the European Research Council, and eSSENCE.

\printbibliography[segment=1]
\end{refsegment}

\end{multicols}

\end{document}